\documentclass[12pt]{article}

\usepackage[english]{babel}
\usepackage[T1]{fontenc}

\usepackage[top=2cm,bottom=2cm,left=2cm,right=2cm,marginparwidth=1.75cm]{geometry}

\usepackage{amsmath, amsthm, amssymb}
\usepackage{graphicx}
\usepackage{natbib}
\usepackage{float}
\usepackage{setspace}
\usepackage[yyyymmdd]{datetime}

\usepackage{xcolor}

\usepackage[colorlinks=true, allcolors=blue]{hyperref}

\newcommand{\bX}{\boldsymbol{X}}
\newcommand{\bx}{\boldsymbol{x}}

\newcommand{\ba}{\boldsymbol{a}}

\newcommand{\bA}{\boldsymbol{A}}
\newcommand{\bw}{\boldsymbol{w}}
\newcommand{\bW}{\boldsymbol{W}}
\newcommand{\bd}{\boldsymbol{d}}

\newcommand{\bv}{\boldsymbol{v}}
\newcommand{\bQ}{\boldsymbol{Q}}
\newcommand{\bq}{\boldsymbol{q}}

\newcommand{\balpha}{\boldsymbol{\alpha}}

\newcommand{\bgamma}{\boldsymbol{\gamma}}

\newcommand{\bSigma}{\boldsymbol{\Sigma}}

\newcommand{\Ber}{\operatorname{Bernoulli}}
\newcommand{\Uni}{\operatorname{Uniform}}
\newcommand{\Mvn}{\operatorname{MVN}}

\newcommand{\bOne}{\boldsymbol{1}}
\newcommand{\bZero}{\boldsymbol{0}}

\newcommand{\mD}{\mathcal{D}}

\title{Survey calibration for causal inference: \\ a~simple method to balance covariate distributions}
\author{Ber{\k e}sewicz Maciej\footnote{
Corresponding author, \url{maciej.beresewicz@ue.poznan.pl}, (1) Poznań University of Economics and Business, Department of Statistics, Al. Niepodległości 10, 61-875 Poznań; (2) Statistical Office in Poznań, Poland. 
This work was financed by the National Science Centre in Poland, OPUS 22 grant no. 2020/39/B/HS4/00941. Codes to reproduce the results are freely available from the github repository: \url{https://github.com/ncn-foreigners/paper-note-quantiles-obs-studies}. An R package that implements the proposed methods is available at \url{https://github.com/ncn-foreigners/jointCalib}. I would like to thank Marcin Szymkowiak, Piotr Chlebicki, Łukasz Chrostowski and Joanna Tyrowicz for their valuable comments on the first draft. I am grateful to Grzegorz Grygiel for his comments and proofreading. I would like to thank Noah Greifer (Harvard University; Institute for Quantitative Social Science) who in a~week after uploading first version of this paper on Arxiv platform (arXiv:2310.11969) implemented the proposed approach in the \texttt{WeightIt} for entropy balancing, non-parametric covariate balancing propensity score, optimization-based  and energy weighting. See more in the development version of the package \url{https://ngreifer.github.io/WeightIt/news/index.html} (\texttt{WeightIt} 0.14.2.9003)
}}

\date{This version: {\today} | First version: \href{https://github.com/ncn-foreigners/paper-note-quantiles-obs-studies/blob/main/paper/2023-beresewicz-causal-balancing.pdf}{2023-10-14}}

\begin{document}
\maketitle

\onehalfspacing

\begin{abstract}
This paper proposes a~simple, yet powerful, method for balancing distributions of covariates for causal inference based on observational studies. The method makes it possible to balance an arbitrary number of quantiles (e.g., medians, quartiles, or deciles) together with means if necessary. The proposed approach is based on the theory of calibration estimators (Deville and Särndal 1992), in particular, calibration estimators for quantiles, proposed by Harms and Duchesne (2006). The method does not require numerical integration, kernel density estimation or assumptions about the distributions. Valid estimates can be obtained by drawing on existing asymptotic theory. An~illustrative  example of the proposed approach is presented for the entropy balancing method and the covariate balancing propensity score method. Results of a~simulation study indicate that the method efficiently estimates average treatment effects on the treated (ATT), the average treatment effect (ATE), the quantile treatment effect on the treated (QTT) and the quantile treatment effect (QTE), especially in the presence of non-linearity and mis-specification of the models. The proposed approach can be further generalized to other designs (e.g. multi-category, continuous) or methods (e.g. synthetic control method). An open source software implementing proposed methods is available.
\end{abstract}



\clearpage

\doublespacing

\section{Introduction}

Recent literature on causal inference for observational studies includes several approaches to balancing whole distributions of covariates rather than moments (e.g. means, variances). In~particular, \cite{hazlett_kernel_2020} proposes kernel entropy balancing (KEB), which consists in making the multivariate density of covariates approximately equal for the treated and control groups when the same choice of kernel is used to estimate these densities. \citet{zhao_covariate_2019} proposes a~covariate balancing rule that modifies the balancing propensity score model by reproducing the kernel Hilbert space and is implemented within the framework of the tailored loss function approach. Finally, \citet{sant2022covariate} proposes an integrated propensity score model, which aims to minimise imbalances in the joint distribution of covariates.

In this paper we propose a~simple method, which consists in balancing control and treatment distributions at specific quantiles along with moments, if needed. Drawing on the theory of calibration estimators in survey sampling \citep{deville1992calibration} and, in particular, on quantile calibration estimators \citep{harms2006calibration}, we develop methods for observational studies. The proposed method involves adding new variables to be balanced based on pre-treatment covariates $\bX$ using a~modified Heaviside function (or its approximation), thus reducing bias and the root mean square error. The procedure does not require a~prior knowledge of the covariate distributions, does not involve integration or tuning parameters such as bandwidths, and is computationally simple as it applies local linear approximation (for example through step-wise regression). 

In this paper, we extend two methods for balancing covariate moments between treatment and control groups, namely entropy balancing proposed by \citet{hainmueller2012entropy}, and the covariate balancing propensity score proposed by \citet{imai2014covariate}. As our method does not add new restrictions on assumptions, nor is it restricted to specific treatment designs, it can be further applied to other methods. For example, our approach has already been extended to multi-category designs and other methods by Noah Greifer (Harvard University; Institute for Quantitative Social Science), who has already implemented our approach in his R package \texttt{WeightIt} (ver. 0.14.2.9003, \citet{weightit}) for binary and multi-category (for selected also continous) treatments using the non-parametric covariate balancing propensity score \citep{fong2018covariate}, inverse probability tilting \citep{graham2012inverse}, optimisation-based weighting \citep{wang2020minimal} and energy balancing weighting \citep{huling2023independence, huling2020energy}. Our approach can be further generalised to other designs (e.g. continuous, longitudinal), assumptions (e.g. treatment is endogenous) or methods (e.g. synthetic control method as an alternative to \citet{gunsilius_distributional_2023}).

The paper is structured as follows. Section \ref{sec-basic} presents the theory underlying calibration estimators for means/totals and quantiles. Section \ref{sec-proposed} describes the proposed approach for the entropy balancing \citep{hainmueller2012entropy} and the covariate balancing propensity score \citep{imai2014covariate}, which we refer to as \textit{distributional entropy balancing} (DEB) and \textit{distributional propensity score} (DPS) methods, respectively. Section \ref{sec-simulation} presents results of two simulation studies aimed at validating the two methods: DEB is used to estimate the average treatment effect on the treated (ATT) and the quantile treatment effect on the treated (QTT), while DPS is used to estimate the average treatment effect (ATE) and the quantile treatment effect (QTE). Section \ref{sec-empirical} presents an empirical example aimed at estimating the ATE and QTE of 401(k) retirement plans on asset accumulation \citep[cf.][]{benjamin2003does} and compares our results with those obtained by \citet{sant2022covariate}. The paper ends with a~conclusion and additional results are presented in the Appendix. 



\section{Theoretical basis from survey sampling}\label{sec-basic}

\subsection{Calibration estimator for a~total}\label{sec-basic-cal}

Let $\bX$ be a~random auxiliary variable and $Y$ be the target random variable of interest. In most applications, the goal is to estimate a~finite population total $~{\tau_{Y}=\sum_{k\in U}Y_{k}}$ or the mean $\bar{\tau}_{Y}=\tau_{Y}/N$ of the variable of interest $Y$, where $U$ is the population of size $N$. The Horvitz-Thompson estimator is a~well-known estimator of a~finite population total, which is expressed as $~{\hat{\tau}_{Y\pi}=\sum_{k=1}^{n}d_{k}Y_{k}=\sum_{k\in s}{d_{k}Y_{k}}}$, where $s$ denotes a~probability sample of size $n$, $d_{k}=1/\pi_{k}$ is a~design weight, and $\pi_{k}$ is the first-order inclusion probability of the $k$-th element of the population $U$. This estimator is unbiased for $\tau_{Y}$ i.e. $E\left(\hat{\tau}_{Y\pi}\right)=\tau_{Y}$. 

Let $\bX_{k}^{\circ}$ be a~$J_{1}$-dimensional vector of auxiliary variables (benchmark variables) for which $~{\tau_{\bX}=\sum_{k\in U}\bX_{k}^{\circ}=\left(\sum_{k\in U}X_{k1},\ldots,\sum_{k\in U}X_{kJ_{1}}\right)^T}$ is assumed to be known. In most cases, in practice the $d_{k}$ weights do not reproduce known population totals for benchmark variables $\bX_{k}^{\circ}$. It means that the resulting estimate $\hat{\tau}_{\bX\pi}=\sum_{k\in s}{d_{k}\bX_{k}^{\circ}}$ is not equal to $\tau_{\bX}$. The main idea of calibration is to look for new calibration weights $w_{k}$ that are as close as possible to original design weights $d_{k}$ and reproduce known population totals $\tau_{\bX}$ exactly. In other words, in order to find new calibration weights $w_{k}$ we have to minimise a~distance function $D\left(\bd,\bv\right)=\sum _{k\in s}d_{k}\hspace{2pt} G\hspace{0pt}\left(\frac{v_{k}}{d_{k}}\right) \to \textrm{min}$ to fulfil calibration equations $\sum_{k\in s}v_{k}\bX_{k}^{\circ} = \sum_{k\in U}\bX_{k}^{\circ}$, where $\bd=\left(d_{1},\ldots,d_{n}\right)^T$, $\bv=\left(v_{1},\ldots,v_{n}\right)^T$ and $G\left(\cdot\right)$ is a~function which must satisfy some regularity conditions: $G\left(\cdot\right)$ is strictly convex and twice continuously differentiable, $G\left(\cdot\right)\geq 0$, $G\left(1\right)=0$, $G'\left(1\right)=0$ and $G''\left(1\right)=1$. Examples of $G\left(\cdot\right)$ functions are given by \citet{deville1992calibration}. For instance, if $G\left(x\right)=\frac{\left(x-1\right)^{2}}{2}$, then using the method of Lagrange multipliers the final calibration weights $w_{k}$ can be expressed as $w_{k}=d_{k}+d_{k}\left(\tau_{\bX}-\hat{\tau}_{\bX\pi}\right)^T\left(\sum_{j\in s}d_{j}\bX_{j}^{\circ}\bX_{j}^{\circ T}\right)^{-1}\bX_{k}^{\circ}$. It is worth adding that in order to avoid negative or large $w_{k}$ weights in the process of minimising the $D\left(\cdot\right)$ function, one can consider some boundary constraints $L\leq \frac{w_{k}}{d_{k}}\leq U$, where $\ 0\leq L\leq 1 \leq U,\  k=1,\ldots,n$. The final calibration estimator of a~population total $\tau_{Y}$ can be expressed as $\hat{\tau}_{Y\bX}=\sum_{k\in s}w_{k}y_{k}$, where $w_{k}$ are calibration weights obtained after selecting a~given $G\left(\cdot\right)$ function. 

Note that the Kullback-Leibler (KL) distance function, given by $~D(d,v) = \sum_k q_k \{v_k \log(\frac{v_k}{d_k}) -v_k + d_k\}$,  reduces to $D(d,v) = \sum_k v_k \log(\frac{v_k}{d_k})$ if $q_k =1$ and $\sum_k v_k = N$. This distance function is known as \textit{entropy divergence} in \citet{hainmueller2012entropy}, \textit{minimum entropy distance} in \citet{deville1992calibration}, \textit{exponential tilting} (EL) in \citet{kim2010calibration} or \textit{generalized exponential tilting} in \citet{wu2016calibration}. After applying the KL distance function one obtains $G(x)=x\log(x) - x + 1$, which is the well known raking ratio method for categorical data and is available in statistical software for sample surveys (e.g. in the \texttt{calibrate} function with \texttt{calfun="raking"} argument in the R \texttt{survey} package; cf. \citet{r-survey}).

\citet{deville1992calibration} and \citet{kim2010calibration} proved that the use of the KL function is asymptotically equivalent to a regression estimator (i.e. using $G(x) = \frac{1}{2}(x-1)^2$ function) for sample surveys (i.e. a calibration estimator can be efficient if $Y_k$ is linearly related to $\bX_k$). This was proved once again by \citet{zhao2016entropy} in the context of observational studies. 

\subsection{Calibration estimator for a~quantile}\label{sec-for-quantiles}

\citet{harms2006calibration} considered a~way of estimating quantiles using the calibration approach, which is very similar to that proposed by \citet{deville1992calibration} for a~finite population total $\tau_{Y}$. By analogy, in their approach it is not necessary to know values for all auxiliary variables for all units in the population. It is enough to know the corresponding quantiles for the benchmark variables. Let us briefly discuss the~problem of finding calibration weights in this setup.  

We want to estimate a~quantile $Q_{Y,\alpha}$ of order $\alpha \in \left(0,1\right)$ of the variable of interest $Y$, which can be expressed as $Q_{Y,\alpha}=\mathrm{inf}\left\{t\left|F_{Y}\left(t\right)\geq \alpha \right.\right\}$, where $F_{Y}\left(t\right)=N^{-1}\sum_{k\in U}H\left(t-Y_{k}\right)$ and the Heaviside function is given by 
\begin{equation}\label{H}
H\left(t-Y_{k}\right)=\left\{ \begin{array}{ll}
1, & \ t \geq Y_{k},\\
0, & \ t<Y_{k}.\\
\end{array} \right.
\end{equation}

We assume that $\bQ_{\bX^*,\alpha}=\left(Q_{X^*_{1},\alpha},\ldots,Q_{X^*_{J_{2}},\alpha}\right)^{T}$ is a~vector of known population quantiles of order $\alpha$ for a~vector of auxiliary variables $\bX_{k}^{*}$, where $\alpha \in \left(0,1\right)$ and $\bX_{k}^{*}$ is a~$J_{2}$-dimensional vector of auxiliary variables. Note that we distinguish $\bX^{\circ}_k$ and $\bX^{*}_k$ auxiliary variables for which we wish to reproduce totals and quantiles respectively. This is because, in general, the numbers $J_{1}$ for $\bX^{\circ}_k$ and $J_{2}$ for $\bX^{*}_k$ may be are different.  It may happen that for a~specific auxiliary variable its population total and the corresponding quantile of order $\alpha$ will be known. However, in most cases, quantiles will be known for continuous auxiliary variables, unlike totals, which will generally be known for categorical variables. In most observational studies unit level data is available for all $\bX$ variables thus possibility of calibration to specific quantiles is easier than for sample surveys.

In order to find new calibration weights $w_{k}$ that reproduce known population quantiles in a~vector $\bQ_{\bX^*,\alpha}$, an interpolated distribution function estimator of $F_{Y}\left(t\right)$ is defined as $
\hat{F}_{Y,cal}(t)=\frac{\sum_{k \in s} w_{k} H_{Y, s}\left(t, Y_{k}\right)}{\sum_{k \in s} w_{k}} 
$, where the Heaviside function in formula (\ref{H}) is replaced by the modified function $H_{Y, s}\left(t, Y_{k}\right)$ given by

\begin{equation}
H_{Y, s}\left(t, Y_{k}\right)=\left\{
\begin{array}{ll}
1, & Y_{k} \leqslant L_{Y, s}(t), \\ 
\beta_{Y, s}\left(t\right), & Y_{k}=U_{Y, s}\left(t\right), \\ 
0, & Y_{k}>U_{Y, s}\left(t\right),
\end{array}\right.
\label{eq-hside}
\end{equation}

\noindent where $L_{Y, s}\left(t\right)=\max \left\{\left\{Y_{k} \mid Y_{k} \leqslant t, k \in s\right\} \cup\{-\infty\}\right\}$, $U_{Y, s}\left(t\right)=\min \left\{\left\{Y_{k} \mid Y_{k}>t, k \in s\right\} \cup\{\infty\}\right\}$ and $\beta_{Y, s}\left(t\right)=\frac{t-L_{Y, s}\left(t\right)}{U_{y, s}\left(t\right)-L_{Y, s}\left(t\right)}$ for $k=1,\ldots,n$, $t \in \mathbb{R}$. A calibration estimator of quantile $Q_{Y,\alpha}$ of order $\alpha$ for variable $Y$ is defined as $\hat{Q}_{Y,cal,\alpha}=\hat{F}_{Y,cal}^{-1}(\alpha)$, where a~vector $\bw=\left(w_{1},\ldots,w_{n}\right)^{T}$ is a~solution of optimization problem $D\left(\bd,\bv\right)=\sum _{k\in s}d_{k}\hspace{2pt} G\hspace{0pt}\left(\frac{v_{k}}{d_{k}}\right) \to \textrm{min}$ subject to the calibration constraints $\sum_{k\in s}v_{k}=N$ and  $\hat{\bQ}_{\bX^*,cal,\alpha}=\left(\hat{Q}_{X^*_{1},cal,\alpha},\ldots,\hat{Q}_{X^*_{J_{2}},cal,\alpha}\right)^{T}=\bQ_{\bX^*,\alpha}$ or equivalently $\hat{F}_{X_{j}^*,cal}\left(Q_{X^*_{j},\alpha}\right)=\alpha$, where $j=1,\ldots,J_{2}$.

As in the previous case, if $G\left(x\right)=\frac{\left(x-1\right)^{2}}{2}$ then using the method of Lagrange multipliers the final calibration weights $w_{k}$ can be expressed as $w_{k}=d_{k}+d_{k}\left(\mathbf{T_{a}}-\sum_{k\in s}{d_{k}\ba_{k}}\right)^{T}\left(\sum_{j\in s}{d_{j}}\ba_{j}\ba_{j}^{T}\right)^{-1}\ba_{k}$, where $\mathbf{T_{a}}=\left(N,\alpha,\ldots,\alpha\right)^{T}$ and the elements of $\ba_{k}=\left(1,a_{k1},\ldots,a_{kJ_{2}}\right)^{T}$ are given by
\begin{equation}
\label{eq-a-vector}
a_{kj}=\left\{\begin{array}{lll} 
N^{-1},& \quad X^*_{kj}\leq L_{X^*_{j},s}\left(Q_{X^*_{j},\alpha}\right),\\
N^{-1}\beta_{X^*_{j},s}\left(Q_{X^*_{j},\alpha}\right), & \quad X^*_{kj}=U_{X^*_{j},s}\left(Q_{X^*_{j},\alpha}\right),\\
0,& \quad X^*_{kj}> U_{X^*_{j},s}\left(Q_{x^*_{j},\alpha}\right),\\
\end{array} \right.
\end{equation}
with $j=1,\ldots,J_{2}$. Alternatively, one can consider the logistic function instead of \eqref{eq-a-vector}

\begin{equation}
a_{kj} = \frac{1}{1+ \exp\left(-2l\left(X^*_{kj}-Q^*_{X^*_j, \alpha}\right)\right)}\frac{1}{N},
\label{eq-logistic-a}
\end{equation}

\noindent where $X^*_{kj}$ is the $k$th row of the auxiliary variable $X^*_j$ ($j=1,...,J_2$), $N$ is the population size, $Q_{X^*_j, \alpha}$ is the known population $\alpha$-th quantile, and $l$ is a~constant set to a~large value (e.g. 1,000). 

In the next sections we describe how this method can be applied to make causal inferences in observational studies. We focus on four causal parameters: ATT, QTT, ATE and QTE.

\section{Proposed approaches}\label{sec-proposed}

\subsection{Setup}

Let us assume that $\mD_k = \{0, 1\}$ is a~treatment indicator variable, sample $s_{0}$ denotes the control group of size $n_0$, $s_{1}$ denotes the treatment group of size $n_1$ and $Y_k$ denotes the variable of interest, where $Y_k(1)$ and $Y_k(0)$ are the potential outcomes for the treatment group and the control group, respectively. The realised outcome is $~{Y_k=\mD_k{Y_k(1)}+(1-\mD_k)Y_k(0)}$ and $\bX_k^{\circ}$ is vector of pre-treatment covariates whose support is denoted as $\mathcal{X}$. Let $p(\bx)=\mathbb{P}(\mD_k=1|\bX_k^{\circ}=\bx)$ be the propensity score where $x \in \mathcal{X}$ and for $\delta = \{0, 1\}$ the distribution and the quantile of the potential outcome $Y_k(\delta)$ is given by $F_{Y_k(\delta)}(y) = \mathbb{P}(Y_k(\delta) \leq y)$ and $q_{Y_k(\delta)}(\alpha) = \mathrm{inf}\left\{t\left|F_{Y_k(\delta)}\left(t\right)\geq \alpha \right.\right\}$.

Let us further assume that the researcher is interested in estimating the average treatment effect on the treated $ATT=\mathbb{E}[Y(1) \mid \mD=1]-\mathbb{E}[Y(0) \mid \mD=1]$, the quantile treatment effect on the treated $QTT(\alpha) = q_{Y(1) \mid D=1}(\alpha)-q_{Y(0) \mid D=1}(\alpha)$, an overall average treatment effect $ATE=\mathbb{E}(Y(1)-Y(0))$ or the quantile treatment effect $QTE(\alpha) = q_{Y(1)}-q_{Y(0)}$.

In this paper we follow a~commonly used identification strategy in policy evaluation \citep[cf][]{rosenbaum1983central, firpo_efficient_2007}: 1) given $\bX_k^{\circ}$, $(Y_k(1),Y_k(0))$ are jointly independent from $\mD_k$ (conditional ignorability), 2) $\forall_{\bx \in \mathcal{X}}\; p(\bx)$ is bounded away from zero and one, 3) uniqueness of quantiles. The identification strategy is the same as that used in the literature mentioned above, since we adopt the same assumptions.

For the ATT and QTT the counterfactual mean and $\alpha$-quantile can be estimated as 

$$
\begin{aligned}
& ATT = \mathbb{E}[\widehat{Y(0)} \mid \mD=1]  =\frac{\sum_{k \in s_0} \omega_{k} Y_{k} }{\sum_{k \in s_0} \omega_{k}},   \\
 & QTT(\alpha) = \mathbb{E}[\widehat{q_{Y(0)}} \mid \mD=1(\alpha)]  = \frac{\sum_{k \in s_0} \omega_{k} H(t - Y_k)}{\sum_{k \in s_0} \omega_{k}},
\end{aligned}
$$

\noindent where $\omega_k$ is a~weight chosen for each control unit.

For ATE one can use the approach suggested by \citet{rosenbaum1987model}, i.e.

\begin{equation}
A T E=\mathbb{E}\left[\left(\frac{\mD}{p(\bX^{\circ})}-\frac{1-\mD}{1-p(\bX^{\circ})}\right) Y\right],
\label{eq-ate}
\end{equation}

\noindent and for QTE with $\delta \in \{0, 1\}$, $F_{Y(\delta)}(y)$ is identified by 

\begin{equation}
F_{Y(\delta)}(y)=\mathbb{E}\left[\frac{1\{\mD=\delta\}}{\delta p(\bX^{\circ})+(1-\delta)(1-p(\bX^{\circ}))} 1\{Y \leqslant y\}\right], 
\label{eq-fyd}
\end{equation}

\noindent where $1\{.\}$ is the indicator function, which means that $QTE(\alpha)$ can also be written as functionals of the observed data \citep[cf.][]{sant2022covariate}.

\subsection{Distributional entropy balancing}

\cite{hainmueller2012entropy} proposed entropy balancing (EB) to reweight the control group to the known characteristics of the treatment group to estimate ATT and QTT. This method can be summarised as follows when only the first moments are constrained

\begin{equation}
\begin{aligned} 
\label{eq-ebal}
\max_{v} H(v)=- & \sum_{k \in s_0} v_{k} \log \left(v_{k} / d_{k}\right) \\
\text { s.t. } & \sum_{k \in s_0} v_{k} X^{\circ}_{kj}=m_{j} \text { for } j \in 1, \ldots, J_1, \\ 
& \sum_{k \in s_0} v_{k}=1 \text { and } v \geq 0 \text { for all } k \in s_0,
\end{aligned}
\end{equation}

\noindent where $v_k$ is defined as previously, $d_k >0$ is the base weight for unit $k$ set to e.g. $d_k=1/n_0$ and $m_j$ is the mean of the $X^{\circ}_j$-th covariate in the treatment group. As in the case of calibration, $\omega_k$ are solutions to \eqref{eq-ebal}.

As we discussed in section \ref{sec-basic-cal} the EB method is a variant of calibration approach where the KL distance function is used. Therefore, we can be simply extended \eqref{eq-ebal} to achieve not only the mean balance but also the distributional balance via quantiles. Instead of using known or estimated population totals $\bQ_{\bX^*,\alpha}$, we can use treatment group quantiles denoted by $\bq_{\bX^*,\alpha} = \left(q_{X^*_{1}, \alpha},\ldots,q_{X^*_{J_{2}},\alpha}\right)^{T}$, where the same $\alpha$ is applied for all $\bX^{*}$ variables and the definition of the vector $\ba_{k}=\left(1,a_{k1},\ldots,a_{kJ_{2}}\right)^{T}$ changes to

\begin{equation}
a_{kj}=\left\{\begin{array}{lll} 
n_1^{-1},& \quad X^*_{kj}\leq L_{X^*_{j},s_0}\left(q_{X^*_{j},\alpha}\right),\\
n_1^{-1}\beta_{X^*_{j},s_0}\left(q_{X^*_{j},\alpha}\right), & \quad X^*_{kj}=U_{X^*_{j},s_0}\left(q_{X^*_{j},\alpha}\right),\\
0,& \quad X^*_{kj}> U_{X^*_{j},s_0}\left(q_{X^*_{j},\alpha}\right),\\
\end{array} \right.
\end{equation}

\noindent with $j=1,\ldots,J_{2}$ where $n_1$ is the size of the treatment group. Alternatively, one can use a~modified \eqref{eq-logistic-a} given by

$$
a_{kj} = \frac{1}{1+ \exp\left(-2l\left(X^*_{kj}-q_{X^*_j, \alpha}\right)\right)}\frac{1}{n_1}.
$$

Our proposal, which leads to distributional entropy balancing (hereinafter DEB), consists of extending the original idea by adding additional constraint(s) on the weights on $\ba_k$, as presented below where the same $\alpha$ is applied for all $\bX^*_j\; j=1,...,J_2$ variables 

$$
\begin{aligned} 
\max _{v} H(v)=- & \sum_{k \in s_0} v_{k} \log \left(v_{k} / d_{k}\right), \\
\text { s.t. } & \sum_{k \in s_0} v_{k} X^*_{k j}=m_{j} \text { for } j \in 1, \ldots, J_1, \\ 
& \sum_{k \in s_0} v_{k} a_{k j}=\frac{\alpha_{j}}{n_1} \text { for } j \in 1, \ldots, J_2, \\ 
& \sum_{k \in s_0} v_{k}=1 \text { and } v \geq 0 \text { for all } k \in s_0.
\end{aligned}
$$

This approach can be easily extended for vector $\balpha$, say quartiles $(0.25, 0.5, 0.75)$ or deciles $(0.1, \ldots, 0.9)$. Our approach is similar to that proposed by \citet{hazlett_kernel_2020}, who extended \eqref{eq-ebal} by replacing the first condition by $\sum_{k \in s_0 } v_{k}\phi\left(X_{k}\right) =\frac{1}{n_{1}} \sum_{k \in s_1} \phi\left(X_{k}\right)$, where $\phi\left(X_{k}\right)$ are the basis functions for the kernel function (in particular the Gaussian kernel). Our approach is simpler as we locally approximate this relationship with a~step-wise (constant) regression because the EB method assumes linear model of $Y_k$ and balancing variables.

\textbf{Remark 1}: \citet{zhao2016entropy} showed that EB method is doubly robust with respect to linear outcome regression and logistic propensity score regression. Including quantiles in the constraints is simply adding new variables to outcome and propensity score regression (as higher order of moments). This leads to step-wise linear/logistic regression, which can approximate non-linear relationships in both models. Effectiveness of this approach will be showed in the simulation study.

\textbf{Remark 2}: Instead of modelling the whole distribution, one can start by adjusting the medians or quartiles. The number of quantiles for $\bX^*_k$ can vary. In the simulation study we show that even a~small number of quantiles significantly improves the estimates, especially in the presence of non-linear relationships.

\textbf{Remark 3}: This approach assumes that the distributions of $\bX_k^*$ between the control and treatment groups have the same support, i.e. it is possible to generate a~vector $\sum_k a_{kj} > 0$.



Our approach can be further applied to hierarchically regularised entropy balancing, as proposed by \citet{xu_hierarchically_2023}, or the empirical likelihood method, as recently discussed by \citet{zhang_calibration_2022}. Development version the \texttt{WeightIt} \citep{weightit} package in R \citep{rcran} currently supports binary and multi-category treatments.

\subsection{Distributional propensity score method}

\citet{imai2014covariate} proposed the covariate balancing propensity score (CBPS) to estimate the \eqref{eq-ate}, where unknown parameters of the propensity score model $\bgamma$ are estimated using the generalized method of moments as

\begin{equation}
\mathbb{E}\left[\left(\frac{\mD}{p\left(\bX^{\circ} ; \bgamma\right)}-\frac{1-\mD}{1-p\left(\bX^{\circ}; \bgamma\right)}\right) f(\bX^{\circ})\right]=\bZero,
\label{eq-cbps}
\end{equation}

\noindent where $p(\dot)$ is the propensity score with unknown vector of parameters $\bgamma \in \mathbb{R}^{J_1}$ (including the intercept). Equation \eqref{eq-cbps} balances means (if $f(\bX^{\circ}=\bX^{\circ})$; or other moments, if specified) of the $\bX^{\circ}$ variables, which may not be sufficient if the variables are highly skewed or we are interested in estimating DTE or QTE. 

We propose a~simple approach based on the specification of moments and $\alpha$-quantiles to be balanced. Instead of using the matrix $\bX^{\circ}$ (including constant), we propose either using $\bX^*$ solely (i.e. balancing only quantiles) as given below or $\bX^{\circ}$ and $\bX^*$ jointly (i.e. balancing both quantiles and moments). Both cases we denote  \textit{distributional propensity score} (DPS) method as in both cases the goal is to balance distribution of $\bX^*$ variables via $\alpha$-quantiles.

To describe the main idea let's focus on the first case where only $\alpha$-quantiles $\bX^*$. In such case the equation \eqref{eq-cbps} is replaced by 

\begin{equation}
\mathbb{E}\left[\left(\frac{\mD}{p\left(\bA ; \bgamma\right)}-\frac{1-\mD}{1-p\left(\bA; \bgamma\right)}\right) f(\bA)\right]=\bZero,
\label{eq-deb}
\end{equation}

\noindent where rows of matrix $\bA$ are given by $\ba_{k}=\left(1,a_{k1},\ldots,a_{kJ_{2}}\right)^{T}$ and elements $\ba_{k}$ for treatment units are given by

\begin{equation}
a_{kj}=\left\{\begin{array}{lll} 
n_1^{-1},& \quad X^*_{kj}\leq L_{X_{j},s_1}\left(q_{X^*_{j,\alpha}}\right),\\
n_1^{-1}\beta_{X_{j}^*,s_1}\left(q_{X^*_{j,\alpha}}\right), & \quad X^*_{kj}=U_{X^*_{j},s_1}\left(q_{X^*_{j,\alpha}}\right),\\
0,& \quad X^*_{kj}> U_{X_{j},s_1}\left(q_{X^*_{j,\alpha}}\right),\\
\end{array} \right.
\end{equation}

\noindent where $n_1$ is the size of the treatment group and $q_{X^*_{j,\alpha}}$ is $\alpha$-quantile for $X_j^*$ for treatment group and for control units

\begin{equation}
a_{kj}=\left\{\begin{array}{lll} 
n_1^{-1},& \quad X^*_{kj}\leq L_{X_{j},s_0}\left(q_{X^*_{j,\alpha}}\right),\\
n_1^{-1}\beta_{X_{j}^*,s_0}\left(q_{X^*_{j,\alpha}}\right), & \quad X^*_{kj}=U_{X^*_{j},s_0}\left(q_{X^*_{j,\alpha}}\right),\\
0,& \quad X^*_{kj}> U_{X_{j},s_0}\left(q_{X^*_{j,\alpha}}\right),\\
\end{array} \right.
\end{equation}

Alternatively, the logistic function \eqref{eq-logistic-a} can be used. Note that the elements of $\bA$ for treatment group will sum up to the selected $\alpha$ orders of the quantiles as $n_1 \to \infty$ (though for small sample sizes, they may not sum up to the specified $\alpha$). As a~result, the propensity score weights balance the $\alpha$ orders of the treatment and control groups and, as shown in \cite{harms2006calibration}, the $\alpha$ quantiles of the selected variables. 

The second case of the DPS method combines both $\bX^{\circ}$ and $\bX^*$ where $\bA$ equation \eqref{eq-deb} is then replaced by $\bX$ as in 

\begin{equation}
\mathbb{E}\left[\left(\frac{\mD}{p\left(\bX; \bgamma\right)}-\frac{1-\mD}{1-p\left(\bX; \bgamma\right)}\right) f(\bX)\right]=\bZero,
\label{eq-deb2}
\end{equation}

\noindent where $\bX = [\bA \; \bX^{\circ}]$ and the intercept in the $\bX^{\circ}$ is removed as $\bA$ already contains it. DPS presented in \eqref{eq-deb2} will preserve not only means but also $\alpha$-quantiles of selected variables if $f(\bX) = \bX$. 

\textbf{Remark 4}. The ATE or QTE estimator based on the CBPS or DPS method is doubly robust if either the outcome or the propensity score is correct. If we use the DPS method presented in \eqref{eq-deb2} we assume that $\mathbb{E}(Y|\bX)$ is approximated by step-wise linear regression with steps created by $\alpha$-quantiles through $\bA$. The same applies to $p(\bX, \bgamma)$, where $\bA$ is used to add the step-wise linear part to e.g. logistic regression. It means that the inclusion of $\alpha$-quantiles makes it possible to approximate non-linear relationships of $Y$ or $\mD$ and $\bX^\circ$ with step-wise linear models. 


As our method simply involves adding new covariates (like adding higher moments), it is possible to use standard procedures to estimate $\bgamma$ parameters with a~just-identified or over-identified set of equations (e.g. generalised method of moments, empirical likelihood). Since we do not change the estimation method itself, any method proposed in the literature can be applied. Furthermore, this matrix can be plugged into a~high-dimensional setting with variable selection as in \citet{ning2020robust}, non-parametric CBPS as in \citet{fong2018covariate} or the improved CBPS proposed by \citet{fan2016improving}. Our approach is not limited to binary cases but can also be used for multi-category or continuous treatments or longitudinal settings. In this way, one can  achieve distribution balancing without developing more sophisticated methods than those proposed by \citet{imai2014covariate}.

The development version of the \texttt{WeightIt} R package currently supports only non-parametric CBPS for binary and multi-category treatments but our approach can be used either by creating matrix $\bA$ independently or by applying our R package \texttt{jointCalib} containing the \texttt{joint\_calib\_cbps} function, which relies on the \texttt{CBPS} package \citep{cbps-pkg}. In the next section we verify the DEB and DPS methods in simulation studies.

\section{Simulation results}\label{sec-simulation}

\subsection{Simulation for the DEB method}

To show the effectiveness of our approach we follow the simulation procedure described by \cite{hainmueller2012entropy}. We generate 6 variables: three ($X_1, X_2$ and $X_3$) from a~multivariate normal distribution $\Mvn(\bZero, \bSigma)$, where

$$
\bSigma = 
\begin{bmatrix}
2 & 1 & -1 \\
1 & 1 & -0.5 \\
-1 & -0.5 & 1 
\end{bmatrix},
$$

\noindent $X_4 \sim \Uni[-3,3]$, $X_5 \sim \chi^2(6)$ and $X_6 \sim \Ber(0.5)$. The treatment and control groups are formed using 

$$
\mD = \bOne[X_1 + 2X_2 - 2X_3 - X_4 -0.5X_5 + X_6 + \epsilon > 0].
$$

We consider three designs: Design 1 (D1): $\epsilon \sim N(0,30)$, Design 2 (D2): $\epsilon \sim N(0,100)$ and Design 3 (D3): $\epsilon \sim \chi^2(5)$ scaled to mean 0.5 and variance 67.6; and three outcome designs:

$$
\begin{array}{l}
Y_1=X_{1}+X_{2}+X_{3}-X_{4}+X_{5}+X_{6}+\eta, \\ 
Y_2=X_{1}+X_{2}+0.2 X_{3} X_{4}-\sqrt{X_{5}}+\eta, \\ Y_3=\left(X_{1}+X_{2}+X_{5}\right)^{2}+\eta,
\end{array}
$$

\noindent where $\eta \sim N(0,1)$. In the simulation study we consider equal sample sizes $n_0=n_1=1000$. As the definition of $\mD$ can lead to unequal sample sizes, we use simple random sampling with replacement from the simulated treatment and control groups to meet the requirement of $n_0=n_1=1000$.

In the simulation study, we use three methods: entropy balancing (EB), kernel entropy balancing (KEB), distributional entropy balancing (DEB) with balancing means and quartiles (denoted as DEB MQ) and DEB with balancing means and deciles (denoted as DEB MD) of $X_1$ to $X_5$. 

Table \ref{tab-deb-results-y3} contains results for ATT and QTT($\alpha$) for $Y_3$ for all designs where $\alpha \in \{0.10, 0.25,0.5,\allowbreak0.75\allowbreak,\allowbreak0.90\}$. Note that we use partially overlapping $\alpha$ for balancing and QTT. Results for $Y_1$ and $Y_2$ are presented in the Appendix~\ref{appen-sec-deb-results}. In all studies we report Monte Carlo  $\text{Bias}=\bar{\hat{\theta}} - \theta$, $\text{Variance}=\frac{1}{R-1}\sum_{r=1}^R\left(\hat{\theta}_r - \bar{\hat{\theta}}\right)^2$ and root mean square error $\text{RMSE}=\sqrt{\text{Bias}^2 + \text{Variance}}$, where $\bar{\hat{\theta}}=\frac{1}{R}\sum_{r=1}^R \hat{\theta}_r$, and $\theta$ is the known effect (i.e. ATT, QTT, ATE or QTE) and $R$ is the number of simulations set to 500.

For all designs, as expected, KEB yields better results in terms of RMSE (mainly thanks to small variance) for $Y_3$, while results for $Y_1$ and $Y_2$ vary. The proposed estimators are better than KEB for $Y_2$ under all three designs, and for $Y_1$ results obtained using DEB are comparable or slightly better than those for KEB. 

Compared with EB, DEB performs better in terms of the RMSE, which is smaller for ATT and QTT(0.25) to QTT(0.90), with a~small increase / and slightly higher / for QTT(0.10) in D1 and D3. The proposed approach improves the estimates of ATT by almost halving the variance of EB for D1 and D2 and significantly reducing the bias for the non-linear case (D3). For D3, an increase in variance is observed for DEB with mean and deciles. DEB MQ and MD are more efficient compared to EB as $\alpha$ increases. 

\clearpage
\begin{table}[ht]
\centering
\small
\caption{Results of simulation for $Y_3$ (strong non-linearity) with $n_0=n_1=1000$ based on $500$ replicates}
\label{tab-deb-results-y3}
\begin{tabular}{llrrrrrr}
  \hline
  Measure & Method & ATT & \multicolumn{5}{c}{QTT} \\ 
         &         &      & 0.10 & 0.25  &  0.50 & 0.75 & 0.90 \\ 
  \hline
  \multicolumn{8}{c}{Design 1 (strong separation, normal errors)} \\
  \hline
  Bias & EB & 0.0044 & 0.0068 & 0.0004 & -0.0031 & -0.0216 & 0.0855 \\ 
       & KEB & 0.0747 & 0.0010 & -0.0108 & 0.0078 & 0.0213 & 0.0675 \\
       & DEB MQ & 0.0141 & 0.0072 & 0.0039 & -0.0021 & -0.0532 & 0.0689 \\ 
       & DEB MD & 0.0098 & 0.0109 & 0.0090 & 0.0052 & -0.0345 & 0.1074 \\ 
  Variance  & EB & 0.4737 & 0.0260 & 0.0366 & 0.1839 & 1.1612 & 6.5252 \\ 
       & KEB & 0.0834 & 0.0270 & 0.0300 & 0.0797 & 0.4980 & 3.4192 \\ 
       & DEB MQ & 0.2999 & 0.0250 & 0.0314 & 0.1326 & 0.9447 & 5.9581 \\ 
       & DEB MD & 0.2513 & 0.0264 & 0.0330 & 0.1233 & 0.8677 & 5.7197 \\ 
  RMSE & EB & 0.6883 & 0.1614 & 0.1914 & 0.4289 & 1.0778 & 2.5559 \\ 
       & KEB  & 0.2983 & 0.1642 & 0.1736 & 0.2825 & 0.7060 & 1.8503 \\ 
       & DEB MQ & 0.5478 & 0.1584 & 0.1773 & 0.3642 & 0.9734 & 2.4419 \\ 
       & DEB MD & 0.5014 & 0.1629 & 0.1819 & 0.3511 & 0.9321 & 2.3940 \\ 
  \hline
  \multicolumn{8}{c}{Design 2 (Weak separation, normal errors)} \\
  \hline
  Bias & EB & 0.0018 & -0.0069 & -0.0040 & -0.0124 & -0.0039 & 0.0618 \\ 
       & KEB & 0.0723 & -0.0123 & -0.0111 & -0.0045 & 0.0524 & 0.0113 \\ 
       & DEB MQ & 0.0006 & -0.0057 & -0.0035 & -0.0080 & -0.0326 & 0.0334 \\ 
       & DEB MD & 0.0041 & -0.0029 & 0.0002 & 0.0025 & -0.0182 & 0.0254 \\ 
  Variance  & EB & 0.5035 & 0.0278 & 0.0355 & 0.2110 & 1.0354 & 4.8602 \\ 
       & KEB & 0.0846 & 0.0290 & 0.0256 & 0.0780 & 0.4865 & 2.6391 \\ 
       & DEB MQ & 0.3527 & 0.0253 & 0.0282 & 0.1320 & 0.8261 & 4.6520 \\ 
       & DEB MD & 0.2514 & 0.0265 & 0.0289 & 0.1222 & 0.6690 & 4.2490 \\ 
  RMSE & EB & 0.7096 & 0.1668 & 0.1885 & 0.4596 & 1.0176 & 2.2055 \\ 
       & KEB & 0.2997 & 0.1708 & 0.1603 & 0.2793 & 0.6995 & 1.6246 \\ 
       & DEB MQ & 0.5939 & 0.1591 & 0.1681 & 0.3634 & 0.9095 & 2.1571 \\ 
       & DEB MD & 0.5014 & 0.1629 & 0.1699 & 0.3496 & 0.8181 & 2.0615 \\ 
  \hline
  \multicolumn{8}{c}{Design 3 (Medium separation, leptokurtic errors)} \\
  \hline
  Bias & EB & 1.1901 & 0.0430 & 0.0977 & 0.4505 & 1.7150 & 3.6745 \\ 
       & KEB & 0.1218 & 0.0834 & 0.1357 & 0.0502 & -0.3744 & 0.8978 \\ 
       & DEB MQ & 0.6893 & 0.0447 & 0.0861 & 0.2746 & 0.8790 & 2.1801 \\ 
       & DEB MD & 0.2402 & 0.0270 & 0.0416 & 0.1144 & 0.3347 & 0.8294 \\ 
  Variance  & EB & 0.5373 & 0.0355 & 0.0447 & 0.2679 & 1.7579 & 12.6890 \\ 
       & KEB & 0.5368 & 0.0353 & 0.0479 & 0.3696 & 2.9304 & 14.3232 \\ 
       & DEB MQ & 0.4965 & 0.0357 & 0.0437 & 0.2719 & 1.9107 & 15.8733 \\ 
       & DEB MD & 0.9612 & 0.0451 & 0.0556 & 0.3337 & 2.3104 & 20.6418 \\ 
  RMSE & EB & 1.3977 & 0.1933 & 0.2330 & 0.6862 & 2.1678 & 5.1177 \\ 
       & KEB & 0.7427 & 0.2055 & 0.2576 & 0.6100 & 1.7523 & 3.8896 \\ 
       & DEB MQ & 0.9857 & 0.1943 & 0.2262 & 0.5893 & 1.6381 & 4.5416 \\ 
       & DEB MD & 1.0094 & 0.2141 & 0.2394 & 0.5889 & 1.5564 & 4.6184 \\ 
   \hline
\end{tabular}
\end{table}

\clearpage

For the non-linear case, DEB MD yields an almost unbiased estimate of the cost of increasing the variance, since the bias for $\alpha=0.90$ in D3 is around 0.83, while for EB it is over 3.6 and the variance is 20.7 and 12.7, respectively. For D2, both the bias and variance decrease compared to their corresponding values for EB, while for D1 only the variance decreases leading to a~decrease in RMSE.

The results suggest that the proposed approach offers more efficient ATT and QTT estimators, especially for non-linear cases in comparison to EB. It should be noted that the DEB approach significantly improves estimates of the upper part of the distribution, which can be beneficial in economic studies. Furthermore, estimation techniques for KEB take a~significant amount of time even for small sample sizes (several minutes), while the proposed approach takes less then a few seconds.

\subsection{Simulation for the DPS method}

In the next simulation, we follow \cite{imai2014covariate} and \cite{sant2022covariate}.  We generate four variables $\bX \sim \Mvn(\bZero, \bSigma)$ where $\bSigma$ is an~$4 \times 4$ identity matrix (for \textit{correctly specified models}). Next, we generate $\bW = (W_1, W_2, W_3, W_4)^T$ with $W_1 =\exp(X_1/2)$, $W_2=X_2/(1+\exp(W_1))$, $W_3=(X_1X_2/25 + 0.6)^3$ and $W_4=(X_2 + X_4 + 20)^T$ (for \textit{mis-specified models} where instead of $\bX$ we observe $\bW$). The true propensity score of the treatment status $\mD$ is given by

$$
p(\bX)=\frac{\exp \left(-X_{1}+0.5 X_{2}-0.25 X_{3}-0.1 X_{4}\right)}{1+\exp \left(-X_{1}+0.5 X_{2}-0.25 X_{3}-0.1 X_{4}\right)},
$$

\noindent and the treatment status $\mD$ is generated $\mD=\bOne\{p(\mathbf{X})>U\}$,  where $U\sim \text{Uniform}[0,1]$. The potential outcomes $Y(1)$ and $Y(0)$ are given by $ Y(1)=210+m(\bX)+\varepsilon(1)$ and $\quad Y(0)=200-m(\bX)+\varepsilon(0)$, where $m(\bX)$ is defined as $m(\bX)=27.4 X_{1}+13.7 X_{2}+13.7 X_{3}+13.7 X_{4}$ and where $\varepsilon(1)$ and $\varepsilon(0)$ are independent $N(0,1)$ random variables. We focus on ATE and QTE($\alpha$) where $\alpha$ is defined as in DEB. The true effect equals 10 for ATE and all QTE. Standard errors were estimated using the same method as in \citet{sant2022covariate} to make the results comparable between the methods. We compare the following approaches:

\begin{itemize}
    \item IPS -- indicator, exponential and projection-based approach proposed by \citet{sant2022covariate} (denoted as IPS (ind), IPS (exp) and IPS (proj)), 
    \item CBPS -- just- and over-identified with balancing means proposed by \citet{imai2014covariate} (denoted as CBPS (j) and CBPS (o)),
    \item DPS -- just- and over-identified with balancing means and quartiles (denoted as DPS (j MQ) and DPS (o MQ) respectively),
    \item DPS -- just- and over-identified with balancing means and deciles (denoted as DPS (j MD) and DPS (o MD) respectively),
    \item DPS -- just- and over-identified with balancing deciles only (denoted as DPS (j D) and DPS (o D) respectively).
\end{itemize}

To compare balance of distributions we use the same metrics as \cite{sant2022covariate}, i.e.

\begin{itemize}
    \item the Cramér-von Mises related statistic (denoted as CVM)
    $$
\text{CVM}(\bgamma)=\sqrt{\frac{1}{n} \sum_{k=1}^{n} \operatorname{DistImb}\left(\tilde{\bX}_{k}, \bgamma\right)^{2}},
$$
    \item the Kolmogorov-Smirnov related statistic (denoted as KS)
    $$
\text{KS}(\bgamma)=\sup_{k: 1, \ldots, n}\left\|\operatorname{DistImb}\left(\tilde{\bX}_{k}, \bgamma\right)\right\|,
$$
\end{itemize}

\noindent where $
\operatorname{DistImb}(\bX, \bgamma)=\mathbb{E}\left[\left(\omega_{1}\left(\mD, \tilde{\bX} ; \bgamma\right)-\omega_{0}\left(\mD, \tilde{\bX} ; \bgamma\right)\right) 1\left\{\tilde{\bX} \leqslant \bx\right\}\right],
$ where $\tilde{\bX}=\bW$ for a~mis-specified model and $\tilde{\bX}=\bX$ for correctly specified model. Note that $\alpha$-quantiles are calculated either on $\bX^*$ or $\bW$ depending which scenario is considered (for the mis-specified model we balance not only on means but also on $\alpha$-quantiles of $\bW$). Weights $\omega_{1}$ and $\omega_{0}$ are defined as

$$
\omega_{1}(\mD, \tilde{\bX} ; \bgamma) =\frac{\mD}{p(\tilde{\bX} ; \bgamma)} / \mathbb{E}\left[\frac{\mD}{p(\tilde{\bX} ; \bgamma)}\right] \; \text{ and } \;
\omega_{0}(\mD, \tilde{\bX} ; \bgamma) =\frac{1-\mD}{1-p(\tilde{\bX} ; \bgamma)} / \mathbb{E}\left[\frac{1-\mD}{1-p(\tilde{\bX} ; \bgamma)}\right].
$$

Table \ref{tabl-sim-2-misspec-results} shows results for a~mis-specified model based on a~Monte Carlo study with 500 replications. Results for correctly specified models are presented in Appendix \ref{app-tab-lin-cbps}. In all cases, DPS performs better than CBPS and IPS in terms of both bias and variance, resulting in a~more efficient estimator. This pattern is observed particularly for the over-identified DPS with decile constraints, since the simulation study only includes continuous variables. The proposed approach leads to nearly unbiased estimates of QTE for all $\alpha$, especially for the upper part of the distribution. 

\begin{table}[ht]
\centering
\scriptsize
\caption{Simulation results for mis-specified model (design 2) based on $500$ replications}
\label{tabl-sim-2-misspec-results}
\begin{tabular}{lrrrrrr}
\hline
  Method & ATE & \multicolumn{5}{c}{QTE} \\ 
           &      & 0.10 & 0.25  &  0.50 & 0.75 & 0.90 \\ 
 \hline
  \multicolumn{7}{c}{Bias} \\
  \hline
  IPS (exp) & 2.0073 & -4.0190 & -2.2247 & 0.7095 & 5.2301 & 10.4740 \\ 
  IPS (ind) & 2.2844 & -3.0931 & -1.3918 & 1.1960 & 5.1096 & 9.4856 \\ 
  IPS (proj) & 0.3018 & -2.6261 & -1.8409 & -0.4423 & 1.9471 & 4.2648 \\ 
  CBPS (j) & 2.6558 & -3.2478 & -1.4447 & 1.4964 & 5.8841 & 10.5782 \\ 
  DPS (j MQ) & 0.6931 & -1.4792 & -0.8639 & -0.0642 & 1.8188 & 4.1355 \\ 
  DPS (j MD) & 0.2633 & -1.2441 & -0.8092 & -0.1933 & 1.1209 & 2.5604 \\ 
  DPS (j D) & -0.4888 & -0.4936 & -0.6743 & -0.5461 & -0.2158 & -0.4357 \\ 
  CBPS (o) & 2.9651 & -2.9019 & -1.1142 & 1.7327 & 6.0674 & 11.2099 \\ 
  DPS (o MQ) & 0.6931 & -1.1227 & -0.6876 & -0.1783 & 1.1199 & 2.9258 \\ 
  DPS (o MD) & 0.2839 & -0.7870 & -0.5274 & -0.0616 & 0.8369 & 1.8841 \\ 
  DPS (o D) & 0.1041 & 0.0239 & -0.1466 & -0.0090 & 0.3576 & 0.3468 \\ 
  \hline
  \multicolumn{7}{c}{Variance} \\
  \hline
  IPS (exp) & 6.7101 & 13.8096 & 9.8939 & 8.0500 & 13.4849 & 38.8827 \\ 
  IPS (ind) & 7.2561 & 15.4085 & 11.1044 & 8.8530 & 14.4145 & 34.6430 \\ 
  IPS (proj) & 5.9519 & 14.6141 & 10.8652 & 7.9917 & 12.8501 & 31.4331 \\ 
  CBPS (j) & 6.5495 & 13.2423 & 9.5448 & 8.4143 & 14.5832 & 35.3577 \\ 
  DPS (j MQ) & 5.9922 & 11.2284 & 8.8770 & 8.5189 & 12.2348 & 24.3055 \\ 
  DPS (j MD) & 6.5039 & 12.2954 & 9.9444 & 9.2286 & 12.8167 & 25.2888 \\ 
  DPS (j D) & 5.7912 & 11.6075 & 9.7880 & 8.7243 & 12.1644 & 22.9389 \\ 
  CBPS (o) & 7.6237 & 12.1805 & 9.4589 & 8.5988 & 15.6431 & 42.3916 \\ 
  DPS (o MQ) & 5.9922 & 10.5597 & 8.4340 & 8.1760 & 11.6608 & 23.6440 \\ 
  DPS (o MD) & 5.4491 & 10.7479 & 8.6797 & 8.0561 & 10.7197 & 19.8878 \\ 
  DPS (o D) & 5.1791 & 10.5077 & 8.5301 & 7.9814 & 10.8403 & 20.0139 \\
  \hline
  \multicolumn{7}{c}{RMSE} \\
  \hline
  IPS (exp) & 3.2771 & 5.4737 & 3.8527 & 2.9246 & 6.3905 & 12.1896 \\ 
  IPS (ind) & 3.5319 & 4.9976 & 3.6113 & 3.2068 & 6.3657 & 11.1633 \\ 
  IPS (proj) & 2.4582 & 4.6380 & 3.7755 & 2.8614 & 4.0794 & 7.0442 \\ 
  CBPS (j) & 3.6882 & 4.8775 & 3.4106 & 3.2640 & 7.0147 & 12.1349 \\ 
  DPS (j MQ) & 2.5441 & 3.6628 & 3.1021 & 2.9194 & 3.9424 & 6.4349 \\ 
  DPS (j MD) & 2.5638 & 3.7207 & 3.2556 & 3.0440 & 3.7514 & 5.6431 \\ 
  DPS (j D) & 2.4556 & 3.4426 & 3.2004 & 3.0037 & 3.4944 & 4.8092 \\ 
  CBPS (o) & 4.0516 & 4.5389 & 3.2711 & 3.4060 & 7.2427 & 12.9635 \\ 
  DPS (o MQ) & 2.5441 & 3.4380 & 2.9844 & 2.8649 & 3.5937 & 5.6749 \\ 
  DPS (o MD) & 2.3515 & 3.3715 & 2.9930 & 2.8390 & 3.3794 & 4.8412 \\ 
  DPS (o D) & 2.2782 & 3.2416 & 2.9243 & 2.8251 & 3.3118 & 4.4871 \\ 
   \hline
\end{tabular}
\end{table}

\begin{table}[ht]
\centering
\caption{Covariate balancing measures for mis-specified model (design 2) based on $500$ replications}
\label{tabl-sim-2-misspec-dists}
\begin{tabular}{lrrrr}
  \hline
  Method & \multicolumn{2}{c}{CVM} & \multicolumn{2}{c}{KS} \\ 
         & Mean & Median & Mean & Median \\ 
  \hline
  IPS (ind)  & 0.28 & 0.26 & 2.29 & 2.22 \\ 
  IPS (exp)  & 0.37 & 0.34 & 2.61 & 2.50 \\ 
  IPS (proj) & 0.75 & 0.53 & 3.33 & 3.08 \\ 
  CBPS (j)   & 0.40 & 0.37 & 2.71 & 2.67 \\ 
  DPS (j MQ) & 0.28 & 0.24 & 2.10 & 2.05 \\ 
  DPS (j MD) & 0.32 & 0.27 & 2.07 & 2.00 \\ 
  DPS (j D)  & 0.31 & 0.26 & 2.11 & 2.06 \\ 
  CBPS (o)   & 0.41 & 0.39 & 2.78 & 2.73 \\ 
  DPS (o MQ) & 0.26 & 0.22 & 2.06 & 2.01 \\ 
  DPS (o MD) & 0.26 & 0.23 & 1.99 & 1.96 \\ 
  DPS (o D)  & 0.26 & 0.22 & 1.99 & 1.94 \\
   \hline
\end{tabular}
\end{table}

Table \ref{tabl-sim-2-misspec-dists} shows a~comparison of the mean and median of the CVM and KS statistics for mis-specified model. In all cases, the proposed method produces more balanced distributions than IPS. Similar pattern is observed for correctly specified model as presented in Appendix with one exception of the IPS method with indicators weights.

\section{Empirical example}\label{sec-empirical}

In this section we study the effect of 401(k) retirement plans on asset accumulation as discussed in \citet{benjamin2003does, belloni2017program, sant2022covariate} and others. We use the dataset sample of 9,910 households from the 1991 SIPP; the treatment assignment follows according to plan eligibility and the outcomes of interest are net financial assets and total wealth. For the purpose of estimation we use the same approach as \citet{sant2022covariate} i.e. for the (instrument) propensity score, we estimate a~logistic specification and use all two-way interactions between income, log-income, age, family size, years of education, dummies for home ownership, marital status, two-earner status, defined benefit pension status, and individual retirement account participation status.

In the study we compare just-identified CBPS, the IPS method involving projection-based weights, as suggested in \citet{sant2022covariate}, and the proposed DPS, in which just-identified CBPS are extended by the inclusion of $\alpha$-quantiles. In our method we balance means and quantiles of income, log-income, age, family size and years of education. For the first three we balance deciles (10\%, 20\%,.., 90\%) and for the last two we balance quartiles and 90\% percentile.

Table \ref{tab-empstudy} shows point estimates and standard errors (in parentheses) for the effect of 401(k) eligibility (participation among compliers) on net financial assets and total wealth. We present estimators for ATE and QTE ($\alpha=0.1,0.25,0.5,0.75,0.90$). We also report two measures of covariate distributional imbalance for each estimator.

\begin{table}[ht!]
\centering
\caption{Effects of 401(k) plan on different measures of wealth}
\label{tab-empstudy}
\begin{tabular}{r|rrr|rrr}
  \hline
 & CBPS & IPS & DPS & CBPS & IPS & DPS \\ 
  \hline
  & \multicolumn{3}{c}{Net Financial Assets} & \multicolumn{3}{c}{Total Wealth} \\
  \hline
  ATE       & 8,190 & 7,784 & 8,014 & 5,997 & 5,395 & 5,518 \\ 
            & (1,150) & (1,605) & (1,086) & (1,811) & (2,792) & (1,756)\\ 
  QTE(0.10) & 1,200 & 1,050 & 1,200 & 400 & 375 & 375 \\ 
            & (254) & (283) & (252) & (558) & (577) & (562) \\ 
  QTE(0.25) & 996 & 996 & 996 & 2,917 & 2,950 & 2,892 \\ 
            & (228) & (231) & (234) & (591) & (617) & (591) \\ 
  QTE(0.50) & 4,200 & 4,300 & 4,136 & 7,419 & 7,419 & 6,943 \\ 
            & (259) & (309) & (259) & (1,111) & (1,157) & (1,114) \\ 
  QTE(0.75) & 12,995 & 12,859 & 12,834 & 8,871 & 8,665 & 8,384 \\ 
            & (922) & (1,025) & (924) & (2,786) & (3,158) & (2,738) \\ 
  QTE(0.90) & 21,053 & 20,899 & 20,980 & 15,979 & 15,504 & 16,091 \\ 
            & (2,247) & (2,782) & (2,265) & (5,829) & (7,818) & (5,697) \\ 
  \hline
  KS & 2.76 & 2.31 & 2.60 & 2.76 & 2.31 & 2.60 \\ 
  CVM & 0.56 & 0.57 & 0.53 & 0.56 & 0.57 & 0.53 \\ 
   \hline
\end{tabular}
\end{table}

The results show that the proposed approach is more efficient than the IPS method. For both outcomes, the standard errors of both ATE and QTE are smaller (with the exception of QTE(0.25) for net financial assets). In particular, the standard errors for total wealth are smaller for the ATE and the upper part of the distribution. The point estimates of the DPS method are close to those of the IPS and CBPS, with the largest difference in QTE(0.50) for total wealth (6,943 for DPS and 7,419 for IPS and CBPS).

When we compare covariate distribution imbalance metrics we get different results for KS and CVM. While the CVM statistic for the proposed method is smaller than that for the IPS and CPS method (0.53 vs 0.57 and 0.56) the KS statistic results for the DPS method are larger than for IPS but smaller than those for CBPS. This may be because the range of propensity scores for the DPS method is between 0.001153 to 0.912971 whereas it is between 0.008307 and 0.868753 for IPS. 

As can be seen, the proposed approach yields more similar results to the method proposed by \citet{sant2022covariate} with lower standard errors. Furthermore, the computational time to fit the parameters based on the IPS method is several hours, while the DPS method takes only a few seconds (on MacBook Air M2 16 GB RAM with 8 cores).

\section{Summary}

In this paper we have proposed a~simple method for balancing distributions based on the theory of calibration estimators for quantiles. The proposed methods are flexible and allow the researcher to balance an arbitrary number of quantiles that may vary according to pre-treatment variables. In particular, if the researcher is interested in estimating a~particular $\alpha$ quantile of treatment effects, they can focus on balancing only these $\alpha$ quantiles for continuous variables. 

Furthermore, the proposed methods perform well for linear and especially for nonlinear and misspecified models. In the two simulation studies, we show that DEB and DPS reduce bias and RMSE for both average and quantile treatment effects. The DEB method is comparable to kernel entropy balancing, but significantly faster and less complicated. DPS outperformed the recently proposed integrated propensity score method. The proposed methods are computationally simple and can be implemented in existing statistical software (e.g. Stata, Python). For the purpose of this study, we developed the \texttt{jointCalib} package, which allows the user to run DEB with different distance functions, such as raking, logit, hyperbolic sinus or empirical likelihood. The DPS method is based on the \texttt{CBPS} package and uses its estimation techniques to balance means and quantiles. For implementation for other methods we suggest to use the \texttt{Weightit} package and methods that allow to specify \texttt{quantile} parameter.

The main limitation of the proposed approach is the uniqueness of the quantiles. For example, one may be interested in balancing the first and second deciles, while these values may be exactly the same in the treatment group (e.g. equal to 0). In such cases, the researcher has to carefully study the distribution of the pre-treatment variables in the control and treatment groups in order to select appropriate quantiles. This problem is also related to the selection of $\alpha$-quantiles, which could be solved with an appropriate penalty, but this requires further research.

Further work may involve adapting this approach to high-dimensional settings and variable selection (which should be selected first: variables or $\alpha$-quantiles?), comparison with double/debiased machine learning framework of \citet{Chernozhukov2018} or synthetic control methods by modifying the set of control variables to account for quantiles rather than means. This approach may be an attractive alternative to \citet{chen_distributional_2020} and \citet{gunsilius_distributional_2023}.

\clearpage

\bibliography{bibliography}
\bibliographystyle{apalike}

\clearpage

\appendix
\setcounter{table}{0}
\renewcommand{\thetable}{A\arabic{table}}

\begin{center}
    \Large
    Appendix for the paper 
    \textit{Survey calibration for causal inference: a~simple method to balance covariate distributions}
\end{center}

\section{Simulation DEB}\label{appen-sec-deb-results}

\begin{table}[ht]
\centering
\scriptsize
\caption{Results of simulation for $Y_1$ (linear) with $n_0=n_1=1000$ based on $500$ replicates}
\label{tab-deb-results-y1}
\begin{tabular}{llrrrrrr}
  \hline
  Measure & Method & ATT & \multicolumn{5}{c}{QTT} \\ 
         &         &      & 0.10 & 0.25  &  0.50 & 0.75 & 0.90 \\ 
  \hline
  \multicolumn{8}{c}{Design 1 (strong separation, normal errors)} \\
  \hline
  Bias & EB & -0.0015 & 0.0029 & 0.0137 & 0.0032 & 0.0049 & -0.0033 \\ 
   & KEB  & 0.0050 & 0.0173 & 0.0188 & 0.0077 & 0.0011 & -0.0044 \\ 
   & DEB MQ & -0.0010 & 0.0060 & 0.0138 & 0.0024 & 0.0026 & 0.0005 \\ 
   & DEB MD & -0.0004 & 0.0104 & 0.0111 & 0.0052 & 0.0067 & 0.0046 \\ 
  Variance & EB & 0.0044 & 0.0617 & 0.0325 & 0.0269 & 0.0378 & 0.0967 \\ 
   & KEB  & 0.0062 & 0.0477 & 0.0289 & 0.0246 & 0.0383 & 0.0830 \\ 
   & DEB MQ & 0.0045 & 0.0581 & 0.0327 & 0.0274 & 0.0384 & 0.0984 \\ 
   & DEB MD & 0.0048 & 0.0587 & 0.0342 & 0.0283 & 0.0402 & 0.1000 \\ 
  RMSE & EB & 0.0664 & 0.2483 & 0.1808 & 0.1642 & 0.1944 & 0.3111 \\ 
   & KEB  & 0.0789 & 0.2191 & 0.1711 & 0.1569 & 0.1958 & 0.2881 \\ 
   & DEB MQ & 0.0669 & 0.2412 & 0.1815 & 0.1654 & 0.1960 & 0.3137 \\ 
   & DEB MD & 0.0696 & 0.2425 & 0.1852 & 0.1684 & 0.2005 & 0.3162 \\
  \hline
  \multicolumn{8}{c}{Design 2 (Weak separation, normal errors)} \\
  \hline
  Bias & EB & -0.0040 & 0.0019 & 0.0021 & -0.0083 & 0.0028 & 0.0021 \\ 
   & KEB & 0.0026 & 0.0152 & 0.0056 & -0.0082 & 0.0029 & 0.0076 \\ 
   & DEB MQ & -0.0037 & 0.0031 & 0.0017 & -0.0078 & 0.0016 & -0.0016 \\ 
   & DEB MD & -0.0035 & 0.0048 & 0.0007 & -0.0089 & 0.0029 & 0.0097 \\ 
  Variance & EB & 0.0040 & 0.0602 & 0.0358 & 0.0260 & 0.0332 & 0.0783 \\ 
   & KEB & 0.0054 & 0.0489 & 0.0288 & 0.0255 & 0.0333 & 0.0622 \\ 
   & DEB MQ & 0.0041 & 0.0579 & 0.0356 & 0.0271 & 0.0347 & 0.0828 \\ 
   & DEB MD & 0.0041 & 0.0626 & 0.0368 & 0.0272 & 0.0356 & 0.0809 \\ 
  RMSE & EB & 0.0631 & 0.2453 & 0.1892 & 0.1614 & 0.1821 & 0.2799 \\ 
   & KEB & 0.0734 & 0.2216 & 0.1699 & 0.1598 & 0.1824 & 0.2494 \\ 
   & DEB MQ & 0.0639 & 0.2407 & 0.1886 & 0.1647 & 0.1862 & 0.2878 \\ 
   & DEB MD & 0.0642 & 0.2502 & 0.1920 & 0.1653 & 0.1886 & 0.2846 \\ 
  \hline
  \multicolumn{8}{c}{Design 3 (Medium separation, leptokurtic errors)} \\
  \hline
  Bias & EB & 0.0066 & -0.2357 & -0.1322 & 0.0030 & 0.1698 & 0.2624 \\ 
   & KEB  & 0.0898 & 0.1145 & 0.0592 & 0.0281 & 0.1019 & 0.1635 \\ 
   & DEB MQ & 0.0053 & -0.2230 & -0.1213 & 0.0075 & 0.1644 & 0.2337 \\ 
   & DEB MD & 0.0062 & -0.2437 & -0.1266 & 0.0134 & 0.1873 & 0.2254 \\ 
  Variance & EB & 0.0081 & 0.0797 & 0.0465 & 0.0400 & 0.0613 & 0.1733 \\ 
   & KEB & 0.0212 & 0.0749 & 0.0536 & 0.0731 & 0.1215 & 0.2628 \\ 
   & DEB MQ & 0.0089 & 0.0887 & 0.0518 & 0.0462 & 0.0695 & 0.1828 \\ 
   & DEB MD & 0.0114 & 0.1369 & 0.0731 & 0.0595 & 0.0824 & 0.2349 \\ 
  RMSE & EB & 0.0901 & 0.3678 & 0.2530 & 0.2000 & 0.3003 & 0.4920 \\ 
   & KEB & 0.1710 & 0.2966 & 0.2389 & 0.2718 & 0.3631 & 0.5381 \\ 
   & DEB MQ & 0.0943 & 0.3721 & 0.2578 & 0.2151 & 0.3107 & 0.4873 \\ 
   & DEB MD & 0.1071 & 0.4430 & 0.2986 & 0.2443 & 0.3428 & 0.5345 \\ 
   \hline
\end{tabular}
\end{table}

\begin{table}[ht]
\centering
\small
\caption{Results of simulation for $Y_2$ (medium non-linearity) with $n_0=n_1=1000$ based on $500$ replicates}
\label{tab-deb-results-y2}
\begin{tabular}{llrrrrrr}
  \hline
  Measure & Method & ATT & \multicolumn{5}{c}{QTT} \\ 
         &         &      & 0.10 & 0.25  &  0.50 & 0.75 & 0.90 \\ 
  \hline
  \multicolumn{8}{c}{Design 1 (strong separation, normal errors)} \\
  \hline
  Bias & EB & -0.0027 & -0.0047 & 0.0003 & 0.0097 & -0.0072 & -0.0075 \\ 
   & KEB & 0.0053 & 0.0122 & 0.0039 & 0.0096 & 0.0013 & -0.0014 \\ 
   & DEB MQ & -0.0016 & 0.0001 & -0.0016 & 0.0099 & -0.0062 & -0.0088 \\ 
   & DEB MD & -0.0014 & -0.0048 & -0.0029 & 0.0101 & -0.0008 & -0.0062 \\ 
  Variance & EB & 0.0049 & 0.0480 & 0.0302 & 0.0213 & 0.0309 & 0.0713 \\ 
   & KEB   & 0.0064 & 0.0377 & 0.0245 & 0.0233 & 0.0283 & 0.0637 \\ 
   & DEB MQ & 0.0049 & 0.0411 & 0.0235 & 0.0200 & 0.0266 & 0.0581 \\ 
   & DEB MD & 0.0055 & 0.0384 & 0.0239 & 0.0210 & 0.0264 & 0.0563 \\ 
  RMSE & EB & 0.0700 & 0.2192 & 0.1739 & 0.1462 & 0.1759 & 0.2672 \\ 
   & KEB & 0.0803 & 0.1946 & 0.1564 & 0.1528 & 0.1683 & 0.2523 \\ 
   & DEB MQ & 0.0702 & 0.2026 & 0.1534 & 0.1416 & 0.1632 & 0.2412 \\ 
   & DEB MD & 0.0740 & 0.1959 & 0.1546 & 0.1453 & 0.1626 & 0.2373 \\  
  \hline
  \multicolumn{8}{c}{Design 2 (Weak separation, normal errors)} \\
  \hline
  Bias & EB & -0.0028 & 0.0047 & 0.0034 & -0.0039 & 0.0009 & 0.0008 \\ 
   & KEB & -0.0027 & 0.0098 & -0.0000 & -0.0066 & -0.0034 & -0.0031 \\ 
   & DEB MQ & -0.0028 & 0.0055 & 0.0014 & -0.0053 & 0.0016 & -0.0025 \\ 
   & DEB MD & -0.0028 & 0.0009 & -0.0004 & -0.0011 & 0.0044 & -0.0004 \\ 
  Variance & EB & 0.0045 & 0.0559 & 0.0314 & 0.0194 & 0.0269 & 0.0598 \\ 
   & KEB  & 0.0053 & 0.0448 & 0.0268 & 0.0187 & 0.0227 & 0.0511 \\ 
   & DEB MQ & 0.0045 & 0.0468 & 0.0263 & 0.0173 & 0.0209 & 0.0532 \\ 
   & DEB MD & 0.0045 & 0.0442 & 0.0254 & 0.0169 & 0.0220 & 0.0472 \\ 
  RMSE & EB & 0.0673 & 0.2364 & 0.1773 & 0.1393 & 0.1641 & 0.2445 \\ 
   & KEB & 0.0726 & 0.2119 & 0.1637 & 0.1370 & 0.1505 & 0.2261 \\ 
   & DEB MQ & 0.0670 & 0.2164 & 0.1621 & 0.1317 & 0.1445 & 0.2306 \\ 
   & DEB MD & 0.0668 & 0.2101 & 0.1595 & 0.1300 & 0.1484 & 0.2171 \\ 
  \hline
  \multicolumn{8}{c}{Design 3 (Medium separation, leptokurtic errors)} \\
  \hline
  Bias & EB & 0.0388 & -0.3582 & -0.2149 & -0.0179 & 0.2204 & 0.5335 \\ 
   & KEB  & 0.1511 & 0.0808 & 0.1121 & 0.0830 & 0.0193 & 0.2120 \\
   & DEB MQ & 0.0504 & -0.2129 & -0.0955 & -0.0033 & 0.1351 & 0.4001 \\ 
   & DEB MD & 0.0697 & -0.1018 & -0.0345 & 0.0093 & 0.1056 & 0.3457 \\ 
  Var & EB & 0.0088 & 0.0553 & 0.0267 & 0.0256 & 0.0505 & 0.1157 \\ 
   & KEB  & 0.0178 & 0.0436 & 0.0257 & 0.0353 & 0.1080 & 0.2385 \\ 
   & DEB MQ & 0.0095 & 0.0465 & 0.0242 & 0.0269 & 0.0524 & 0.1252 \\ 
   & DEB MD & 0.0150 & 0.0514 & 0.0361 & 0.0387 & 0.0712 & 0.1394 \\ 
  RMSE & EB & 0.1014 & 0.4285 & 0.2700 & 0.1611 & 0.3148 & 0.6327 \\ 
   & KEB & 0.2016 & 0.2239 & 0.1956 & 0.2053 & 0.3292 & 0.5324 \\ 
   & DEB MQ & 0.1099 & 0.3030 & 0.1825 & 0.1642 & 0.2658 & 0.5341 \\ 
   & DEB MD & 0.1410 & 0.2485 & 0.1931 & 0.1970 & 0.2870 & 0.5088 \\  
   \hline
\end{tabular}
\end{table}

\clearpage

\section{Simulation for propensity score}

\begin{table}[ht]
\centering
\scriptsize
\caption{Simulation results for correctly specified model (design 1) based on $500$ replications}
\label{app-tab-lin-cbps}
\begin{tabular}{lrrrrrr}
\hline
  Method & ATE & \multicolumn{5}{c}{QTE} \\ 
           &      & 0.10 & 0.25  &  0.50 & 0.75 & 0.90 \\ 
 \hline
  \multicolumn{7}{c}{Bias} \\
  \hline
  IPS (exp) & -0.0378 & 0.0279 & -0.0371 & -0.2570 & 0.0195 & 0.0706 \\ 
  IPS (ind) & 0.3536 & 0.1618 & 0.1559 & 0.1256 & 0.5079 & 0.7946 \\ 
  IPS (proj) & -0.0390 & 0.0177 & -0.0340 & -0.2510 & 0.0144 & 0.0863 \\ 
  CBPS (j) & -0.1158 & -0.0334 & -0.0620 & -0.3772 & -0.0748 & -0.0619 \\ 
  DPS (j MQ) & -0.0485 & -0.0204 & -0.0619 & -0.2350 & 0.0113 & 0.2196 \\ 
  DPS (j MD) & -0.0208 & -0.0165 & -0.0654 & -0.1872 & 0.0307 & 0.3950 \\ 
  DPS (j D) & -0.4777 & 1.0186 & 0.4195 & -0.3893 & -1.1203 & -2.1375 \\ 
  CBPS (o) & -0.1303 & -0.0628 & -0.1349 & -0.3353 & -0.0715 & -0.0208 \\ 
  DPS (o MQ) & -0.0485 & 0.0395 & -0.0649 & -0.3007 & -0.1279 & -0.0503 \\ 
  DPS (o MD) & -0.0220 & 0.0056 & -0.0843 & -0.2132 & 0.0577 & 0.3011 \\ 
  DPS (o D) & 0.0084 & 0.9148 & 0.4509 & -0.0334 & -0.2882 & -0.8203 \\
  \hline
  \multicolumn{7}{c}{Variance} \\
  \hline
  IPS (exp) & 6.8488 & 12.8167 & 9.5348 & 8.5481 & 14.6863 & 36.0200 \\ 
  IPS (ind) & 6.4538 & 13.5540 & 10.8244 & 9.1452 & 14.1257 & 31.2927 \\ 
  IPS (proj) & 6.6207 & 12.8720 & 9.5067 & 8.4839 & 14.2258 & 34.4839 \\ 
  CBPS (j) & 7.7502 & 11.2431 & 9.0807 & 9.7304 & 17.0821 & 39.5007 \\ 
  DPS (j MQ) & 7.1257 & 11.1017 & 9.4488 & 9.2409 & 14.6424 & 35.2882 \\ 
  DPS (j MD) & 7.0747 & 12.3001 & 9.9328 & 9.6852 & 15.7340 & 35.5981 \\ 
  DPS (j D) & 6.1781 & 11.1420 & 9.5545 & 9.3014 & 14.5493 & 29.7210 \\ 
  CBPS (o) & 7.6881 & 11.2360 & 9.0135 & 9.6730 & 16.5078 & 37.8701 \\ 
  DPS (o MQ) & 7.1257 & 11.0554 & 9.0681 & 8.6422 & 13.2662 & 28.7042 \\ 
  DPS (o MD) & 5.8066 & 11.2075 & 9.0012 & 8.5178 & 12.4013 & 24.7986 \\ 
  DPS (o D) & 5.4536 & 10.9623 & 8.9961 & 8.4550 & 12.2454 & 23.0597 \\ 
  \hline
  \multicolumn{7}{c}{RMSE} \\
  \hline
  IPS (exp) & 2.6173 & 3.5802 & 3.0881 & 2.9350 & 3.8323 & 6.0021 \\ 
  IPS (ind) & 2.5649 & 3.6851 & 3.2937 & 3.0267 & 3.7926 & 5.6501 \\ 
  IPS (proj) & 2.5734 & 3.5878 & 3.0835 & 2.9235 & 3.7717 & 5.8729 \\ 
  CBPS (j) & 2.7863 & 3.3532 & 3.0141 & 3.1421 & 4.1337 & 6.2853 \\ 
  DPS (j MQ) & 2.6698 & 3.3320 & 3.0745 & 3.0490 & 3.8266 & 5.9444 \\ 
  DPS (j MD) & 2.6599 & 3.5072 & 3.1523 & 3.1177 & 3.9667 & 5.9795 \\ 
  DPS (j D) & 2.5311 & 3.4899 & 3.1194 & 3.0746 & 3.9755 & 5.8557 \\ 
  CBPS (o) & 2.7758 & 3.3526 & 3.0053 & 3.1282 & 4.0636 & 6.1539 \\ 
  DPS (o MQ) & 2.6698 & 3.3252 & 3.0120 & 2.9551 & 3.6445 & 5.3579 \\ 
  DPS (o MD) & 2.4098 & 3.3478 & 3.0014 & 2.9263 & 3.5220 & 4.9889 \\ 
  DPS (o D) & 2.3353 & 3.4350 & 3.0330 & 2.9079 & 3.5112 & 4.8716 \\ 
   \hline
\end{tabular}
\end{table}

\begin{table}[ht]
\centering
\caption{Covariate balancing measures for correctly specified model (design 1) based on $500$ replications}
\begin{tabular}{lrrrr}
  \hline
  Method & \multicolumn{2}{c}{CVM} & \multicolumn{2}{c}{KS} \\ 
         & Mean & Median & Mean & Median \\ 
  \hline
  IPS (ind) & 0.15 & 0.15 & 1.95 & 1.89 \\ 
  IPS (exp) & 0.20 & 0.19 & 2.09 & 2.02 \\ 
  IPS (proj) & 0.20 & 0.19 & 2.08 & 2.02 \\ 
  CBPS (j) & 0.22 & 0.20 & 2.15 & 2.10 \\ 
  DPS (j MQ) & 0.19 & 0.17 & 1.96 & 1.92 \\ 
  DPS (j MD) & 0.18 & 0.17 & 1.87 & 1.83 \\ 
  DPS (j D) & 0.18 & 0.17 & 1.92 & 1.88 \\ 
  CBPS (o) & 0.21 & 0.20 & 2.15 & 2.08 \\ 
  DPS (o MQ) & 0.20 & 0.18 & 2.06 & 1.98 \\ 
  DPS (o MD) & 0.22 & 0.21 & 2.18 & 2.16 \\ 
  DPS (o D) & 0.23 & 0.21 & 2.24 & 2.22 \\ 
   \hline
\end{tabular}
\end{table}

\end{document}